\documentstyle[12pt,a4]{article}

\def\be{\begin{equation}}
\def\ee{\end{equation}}
\def\bea{\begin{eqnarray}}
\def\eea{\end{eqnarray}}
\def\<{\langle}
\def\>{\rangle}

\newcommand{\rf}[1]{(\ref{#1})}
\newcommand{\CR}{\nonumber \\}

\begin{document}

\begin{titlepage}
\title{
\hfill\parbox{4cm}
{\normalsize KUNS-1484\\HE(TH)~97/20\\{\tt hep-th/9712180}}\\
\vspace{1cm}
M-theory description of BPS string \\ in 7-brane background }
\author{
Isao {\sc Kishimoto}\thanks{{\tt ikishimo@gauge.scphys.kyoto-u.ac.jp}}
and
Naoki {\sc Sasakura}\thanks{{\tt sasakura@gauge.scphys.kyoto-u.ac.jp}}
\\[7pt]
{\it Department of Physics, Kyoto University, Kyoto 606-01, Japan}
}
\date{\normalsize December, 1997}
\maketitle
\thispagestyle{empty}

\begin{abstract}
\normalsize
We discuss  the BPS configurations of IIB strings 
in the 7-brane background from M-theory viewpoint. 
We first obtain the hyperk\"ahler geometry background 
of M-theory expected from 
the 7-brane solutions of the type IIB supergravity.
We choose the appropriate complex structures of the background geometries and 
embed a membrane of M-theory holomorphically to obtain a BPS string
configuration.
The recently discussed BPS string configurations such as 3-string
junctions and string networks in the flat background
are generalized to the cases with the 7-brane backgrounds.
The property of the BPS string configurations in the 7-brane 
backgrounds is in agreement with the previously known results from the 
IIB string viewpoint.
\end{abstract}

\end{titlepage}

\renewcommand{\thefootnote}{\fnsymbol{footnote}}
\setcounter{footnote}{1}

The analysis of BPS states in a brane configuration 
is important in identifying the low energy effective field theory
associated to the brane configuration and investigating its  
non-perturbative dynamics.
An open string connecting these branes can be such BPS states.
In the references \cite{JOH,ZWI}, they analyzed the structure of the BPS
states in the 7-brane background of F-theory and identified the 
open string states
corresponding to the gauge bosons of exceptional gauge groups.
The open string BPS configurations with 3-string
junctions were essential in the analysis of \cite{ZWI}.
They assumed that these configurations are BPS under certain conditions, and 
argued that a 3-string junction arises naturally when a string
crosses a 7-brane in course of changing the moduli parameters, 
since this effect is actually U-dual to the 
Hanany-Witten effect \cite{HANWIT}.
The BPS nature of the 3-string junction was originally conjectured 
in \cite{SHW}, and have recently been proven in \cite{DAS,SENONE,
REY,KROLEE,MAT}.
Joining many of these 3-junctions, BPS string network was 
constructed in \cite{SHW,SENONE,KROLEE}.
Similar fivebrane junctions and networks are discussed in the 
literatures \cite{KOL}.

The BPS conditions of a 3-string junction are given as follows. 
Consider a junction of three 
$(p_i,q_i)$ strings ($i=1,2,3$) 
stretching in $\vec{n}_i$ directions in $x_8$-$x_9$ plane 
($|n_i|=1$). 
One condition is that the charges of the strings should be conserved:
\be
\sum_{i=1}^3 p_i=\sum_{i=1}^3 q_i= 0.
\label{chacon}
\ee
The other is that the forces from each string tensions should be balanced:
\be
\sum_{i=1}^3 |p_i +q_i \tau | \vec{n}_i=0, 
\quad \tau=\chi+i e^{-\phi},
\label{balcon}
\ee
where $\chi$ and $\phi$ are the axion and the dilaton fields of the IIB 
string theory, respectively.

In \cite{KROLEE,MAT}, the BPS conditions \rf{chacon}, \rf{balcon} are obtained 
from M-theory viewpoint.
In M-theory, a $(p,q)$ string is a membrane winding $q$ and $p$
times in $x_7$ and $x_{10}$ directions, respectively \cite{SHW}.
The BPS condition is simply that the embedding of the membrane
is a holomorphic map \cite{KROLEE,MAT,BECBEC}.
Since the number of the winding directions of a membrane describing a 
string is one, 
the appropriate complex structure of the target space $z_1$ and $z_2$
must contain each of the coordinates of the compact 
directions $x_7$ and $x_{10}$, respectively.
Taking the complex structure of the flat target space as 
$z_1=x_8+ix_7,\ z_2=x_9+ix_{10}$,
a BPS string configuration is just given by an appropriate holomorphic 
equation $f(e^{z_1},e^{z_2})=0$. 
The authors of \cite{KROLEE,MAT} actually discussed the BPS conditions
\rf{chacon}, \rf{balcon} from the holomorphic equation describing a
BPS 3-junction in a flat background.
 
In this paper we will extend the analysis in M-theory to the case 
with 7-brane backgrounds.
This extension would be interesting
for further understanding of the BPS states and
the generation process of a 3-string junction in 7-brane backgrounds.
We will first obtain the geometrical backgrounds of M-theory from the 
7-brane backgrounds of the IIB theory proposed in
\cite{GRESHA,GIBGRE,BERROO}.
The geometry is hyperk\"ahler.
The complex structure obtained naturally in the process will not be
appropriate to describe a $(p,q)$ string. 
Thus we will choose another one.
Once we obtain the appropriate one, a holomorphic curve
will generally give a BPS string configuration \cite{KROLEE,MAT,BECBEC}.

The type IIB SUGRA action of the metric, the
axion and the dilaton field is given in the Einstein frame by
\bea
S &=& {1\over 2{\kappa}^2} 
\int d^{10} { x} \sqrt{-g} \left(R - {1\over 2} g^{\mu \nu}
{{\partial_{\mu}\tau \partial_{\nu}{\bar \tau}}\over \tau_2^2}\right), \CR
\tau&=&\chi + i e^{-\phi}=\tau_1 + i\tau_2.
\label{suglag}
\eea
This action has supersymmetric solutions \cite{GRESHA,GIBGRE}
under the following ansatz for the metric:
\bea 
ds_{{\rm IIB(E)}}^2 &=& -dx_0^2 + dx_1^2 + \dots +dx_7^2 + e^{\Phi}dz
d{\bar z}, 
\CR
z&=&x_8 + i x_9,\ \ \Phi=\Phi(z,{\bar z}).
\label{ansmet}
\eea
The field $\tau$ is taken to be a holomorphic function $\tau(z)$ 
to preserve half the supersymmetry and 
the field $\Phi$ in \rf{ansmet} is given by 
\be
\Phi = \log \tau_2 + F(z) + {\bar F(\bar z)} = - \phi + F(z)
 + {\bar F(\bar z)}
\label{phival}
\ee
with a holomorphic function $F(z)$.
The field $\tau$ will become the moduli parameter of the 
compactification torus in the M-theory description of the type IIB
string theory.
Hence $\tau$ may have a $SL(2,Z)$ holonomy around a singularity.
Thus the $z$-dependence of $\tau$ can be given by
\be
j(\tau(z)) = {P(z)\over Q(z)},
\label{eqn:jPQ}
\ee
where
$P(z)$ and $Q(z)$ are polynomials in $z$ with no common factors, and the 
function $j(\tau)$ is the $j$-function, a holomorphic one-to-one map 
from the moduli space of a torus 
to the complex plane, defined explicitly by
\bea
j(\tau) &=& { ( \theta_2(\tau)^8 + \theta_3(\tau)^8 +
  \theta_4(\tau)^8)^3 \over \eta(\tau)^{24}}, \CR
\eta(\tau)&=& e^{{\pi i \tau}\over 12} \prod_{n=1}^{\infty}(1-e^{2 n
  \pi i \tau}),
\label{eqn:eta}
\eea
using Jacobi's theta functions and Dedekind's function.
In case ${\rm deg}(P)\leq{\rm deg}(Q)$ and the polynomial $Q(z)$ is given by
\be
Q(z) = \prod_{i=1}^{N} (z-z_i),
\label{forofq}
\ee
the function $F(z)$ is determined by the requirement that it should be
regular at $z_i$ and that the Einstein metric \rf{phival} should be invariant 
under the $SL(2,Z)$ symmetry of \rf{suglag}:
\be
e^{F(z)} = \eta (\tau)^2 \prod_{i=1}^{N} (z-z_i)^{-{1\over 12}}.
\label{eqn:eF}
\ee

In the IIB string frame, the 7-brane background \rf{ansmet} is given by
\bea
ds_{{\rm IIB(str)}}^2 &=& e^{{1\over 2}\phi} ds_{{\rm IIB(E)}}^2 
= e^{{1\over 2}\phi}( -dx_0^2 + dx_1^2 + \dots +dx_7^2)
+ e^{-{1\over 2}\phi + F + {\bar F}}dz d{\bar z},\CR
\tau&=&\tau(z).
\label{baccom}
\eea
Now let us compactify the $x_7$ direction: $x_7=R_B{\b {\it x}_7},\  {\b {\it x}_7}\sim {\b {\it x}_7}+2\pi$.  
After the T-duality transformation \cite{BHO} in this direction, we obtain the
following IIA background:
\bea 
ds_{{\rm IIA(str)}}^2 &=& e^{{1\over 2}\phi}( -dx_0^2 + dx_1^2 + \dots +
dx_6^2)+ e^{-{1\over 2}\phi}R_B^{-2} d{\b {\it x}_7}^2 
+ e^{-{1\over 2}\phi + F + {\bar F}}dz d{\bar z},\CR
\phi_A &=& {3\over 4}\phi - \log R_B, \CR
A^{(1)} &=& \chi d{\b {\it x}_7} ,
\eea
where $\phi_A$ and $A^{(1)}$ are the NS-NS scalar and the R-R 1-form
field of the IIA string theory, respectively.
Decompactifying an 11th direction $x_{10}$ $(x_{10}\sim x_{10}
+2\pi)$ \cite{WIT,BHO}, the M-theory background geometry is obtained as
\bea
ds_{{\rm 11D}}^2 &=& R_B^{2\over 3}(- dx_0^2 + dx_1^2 + \dots +dx_6^2 + ds_4^2),\CR
ds_4^2 &=&  e^{-\phi + F + {\bar F}}dz d{\bar z} + e^{\phi}
R_B^{-2}|dx_{10} + \tau d{\b {\it x}_7} |^2 .
\label{elemet}
\eea
The area of the compactification torus is $(2\pi)^2R_B^{-4/3}$, where 
the decompactification limit of the IIB string theory
is given by $R_B \rightarrow \infty$.
 
Viewing \rf{elemet}, the natural choice of complex structure would be
$z$ and $v=x_{10}+\tau {\b {\it x}_7}$, and the metric is now
\be
ds_4^2 =  e^{-\phi + F + {\bar F}}dz d{\bar z} + e^{\phi}
R_B^{-2}\left| dv - {v-{\bar v}\over {\tau - {\bar \tau}}}\partial \tau dz
\right| ^2 .
\label{mettwo}
\ee
One can easily see that this is in fact K\"ahler with this choice of complex 
structure by showing $dK=0$, where $K$ is the K\"ahler form associated
to it.
This 4-dimensional metric is the same Ricci-flat K\"ahler metric which 
appeared as the geometry of the cosmic string configuration discussed
in \cite{GRESHA}.
A 4-dimensional Ricci-flat K\"ahler metric is 
hyperk\"ahler\footnote{For example, see section 2.2 of \cite{ASP}.}, and 
this is consistent with that the geometry  should preserve 
half the supersymmetry of 11D SUGRA.

The above complex structure is not appropriate to describe a $(p,q)$
string by a holomorphic curve, because the coordinates of the
compactification torus are parameterized by one complex variable $v$.
Since the metric is hyperk\"ahler, we can choose another complex
structure. We would like to find a complex structure in the form
$X = f(z,{\bar z}) + i R_B^{-1}x_{10} ,\  
Y = g(z,{\bar z}) + i R_B^{-1}{\b{\it x}_7}$. 
One can easily find a complex structure in this form by using the fact 
that
the K\"ahler form $K'$ of another complex structure is related to the nowhere
vanishing holomorphic 2-form $\Omega$ of the original complex structure:
\be
K' = c {\rm Re}( \Omega ),\ \ \Omega = e^{F + i \theta}dz \wedge dv
\label{eqn:K'}
\ee
with an arbitrary constant $\theta$ and a real constant normalization
$c$. After a short calculation, we obtain
\bea
X &=& {\rm Im}\left(\int^z \tau e^{F + i \theta}dz\right)+
i R_B ^{-1} x_{10},\CR
Y &=&  - {\rm Im}\left( \int^z e^{F + i \theta}dz\right)+
i R_B ^{-1} {\b {\it x}_7} .
\label{eqn:XY}
\eea
Here the constant normalization of the real parts of $X$ and $Y$ was 
fixed by demanding that the metric \rf{mettwo} be expressed with $X$
and $Y$ in a K\"ahler form. With this complex structure, 
the K\"ahler metric \rf{mettwo} is  
\be
ds_4^2 = \tau_2^{-1}(|dX|^2 + |\tau |^2 |dY|^2 + \tau_1 (dX d{\bar Y}
+dY d{\bar X})) .
\label{mettor}
\ee
This metric looks just like that on the compactification torus
but here the moduli parameter $\tau$ depends non-trivially on $X$ and $Y$.

Once we have obtained the appropriate complex structure, a
holomorphic curve will generally give a BPS string 
configuration including 3-string junctions and 
string networks \cite{KROLEE,MAT}.
Since a $(p,q)$ string is a membrane winding $p$ and $q$ 
times in $x_{10}$ and $x_7$ directions, respectively \cite{SHW},
the curve for a $(p,q)$ string is given by \cite{KROLEE,KOL}  
\be
s^q t^{-p} = {\rm constant},
\label{eqn:st}
\ee    
where $s=\exp(R_BX)$ and $t=\exp(R_BY)$. This is 
\bea
qx_{10} - p {\b {\it x}_7} &=& {\rm constant}, \CR
R_B {\rm Im} \biggl(  e^{i \theta} \int^z h_{p,q}(z)dz \biggr) &=& 
{\rm constant },
\label{mgeo}
\eea
where
\be
h_{p,q}(z) = (p+ q \tau(z) )\eta(\tau(z))^2 \prod_{i=1}^{N}
(z-z_i)^{-{1\over 12}}.
\label{hpq}
\ee

The second equation of \rf{mgeo} describing the path in the
non-compact space is exactly what is expected from 
the IIB string picture. In the IIB picture, the mass of a string is 
obtained by integrating the line elements of the path of the string 
multiplied by the string tension. In the 7-brane background, the mass
of a $(p,q)$ string along a path $C$ is given by \cite{SENTWO}
\bea
m_{p,q}&=&\int_C |d\omega_{p,q}|, \CR
d\omega_{p,q}&=&h_{p,q}(z)dz.
\eea
This shows that a path with a minimal mass is given by a straight line
in the new coordinate defined by $\omega_{p,q}(z)=\int^z h_{p,q}(z)
dz$. This is exactly what is meant by the second equation of \rf{mgeo}.

Another 7-brane solution, which preserves half the supersymmetry and is
T-dual to the 8-brane of the IIA string theory,
was proposed in the literature \cite{BERROO}.
This circularly symmetric solution is given in the IIB string frame by
\bea
ds_{{\rm IIB(str)}}^2&=&H^{-{1\over 2}}(r)(-dx_0^2+dx_1^2+\dots+dx_7^2)+
H^{1\over 2}(r)(dr^2+d{\b {\it x}}^2),\CR
\tau&=&\pm H'(r){\b {\it x}}+i H(r),\CR
H''(r)&=&0,
\eea
where $r$ and $\b{\it x}$ are a radial and an angular coordinate 
${\b {\it x}} \sim {\b {\it x}} +1$, respectively, and the prime symbol $'$ denotes
taking derivative of $r$.
Taking the simplest solution $ H(r)= \pm {\tilde m} (r-r_0) $ 
and choosing the complex structure $z={\tilde m}({\b {\it x}} \pm i
(r-r_0))$, the 7-brane background is given by 
\bea
ds_{{\rm IIB(str)}}^2&=&H^{-{1\over 2}}(z,{\bar z})(-dx_0^2+ dx_1^2 + \dots
+dx_7^2)+H^{1\over 2}(z,{\bar z}){\tilde m}^{-2}dz d{\bar z},\CR
\tau(z) &=& z.
\eea
Since this metric can be obtained just by the substitution 
$e^{-\phi} = H,\ \ e^F={\tilde m}^{-1}$ in \rf{baccom}, 
the correspondent to \rf{hpq} is obtained as 
\be
h^{{\rm cir.}}_{p,q}(z)=\tilde m ^{-1}(p+qz).
\ee
Hence, in this circularly symmetric 7-brane background, the
second equation of \rf{mgeo} gives 
the path of a $(p,q)$ string in the $z$-plane (upper half plane 
because ${\rm Im}(\tau)>0$) by 
\be
{\rm Im}\left(e^{i \theta}{\tilde m}^{-1}\left(
p z+q {1\over 2}z^2\right)\right)={\rm constant}.
\ee
For the general case of $q\ne 0$, the path is hyperbolic in the $z$-plane, 
while it is a straight line in case $q=0$, showing an interesting
charge dependence of the interaction between the 7-brane and a $(p,q)$ 
string.
 
Now we will discuss the transformation property of the complex
structure \rf{eqn:XY} under the $SL(2,Z)$ transformation 
in case that $\tau(z)$ is given by (\ref{eqn:jPQ}) and \rf{forofq}
with ${\rm deg}(P)\leq{\rm deg}(Q)$. 
The parameter $\tau(z)$ given by \rf{eqn:jPQ}
has in general a holonomy around a singularity given by 
\be
\tau' = {{a \tau + b}\over {c \tau +d}}, \ \ 
\left(
\begin{array}{cc}
a & b \\
c & d
\end{array}
\right) \in SL(2,Z).
\label{slttp}
\ee
Since the moduli parameter $\tau'$ represents the same torus
as that represented by $\tau$, we may introduce a branch cut and define a 
transformation property between the two different descriptions on the 
branch cut.
In the case of (\ref{eqn:jPQ}), the singularities come from the
singular points of $\tau$, i.e. 
$\tau=i\infty$ (logarithmic  singularity), 
$\tau=i$ ($Z_2$ singularity), and $\tau=e^{{2 \pi i}\over  3} $($Z_3$
singularity) in the fundamental region, 
and any kinds of cuts may in general appear in the $z$-plane.

As for the imaginary part of the complex structure \rf{eqn:XY},
one can easily see that the coordinates of the torus should be related
in the following way under the transformation \rf{slttp}:
\be
(d\b{\it x}_7' , dx_{10}') = 
(d\b{\it x}_7,dx_{10})
\left(
\begin{array}{cc}
a & b \\
c & d
\end{array}
\right)^{-1}.
\label{eqn:x_10}
\ee
As for the real part of \rf{eqn:XY}, we begin with the 
property of Dedekind's function under the $T$ and $S$ transformations
of the moduli parameter $\tau$:
\bea
T&:&\eta(\tau+1) = e^{{\pi i}\over 12}\eta(\tau),\CR 
S&:&\eta(-1/ \tau) = (-i \tau)^{1\over 2}\eta(\tau).
\label{phase}
\eea
On a branch cut connected to the $\tau=i\infty$ singularity at $z=z_i$, 
the associated $SL(2,Z)$ transformation is the $T$-transformation, so
the phase in \rf{phase} is canceled by the phase from
$(z-z_i)^{-\frac1{12}}$ in \rf{eqn:eF}.
On the other hand, 
on a cut connected to a $Z_2$ or a $Z_3$ singularity, a phase factor
may remain in \rf{eqn:eF}.
We have also a one parameter freedom $\theta$ in the
choice of the complex structure \rf{eqn:K'}.
It turns out that we can introduce the transformation rule of $\theta$ as 
\be
T:\theta'= \theta ,\ \ 
S:\theta'= \theta + {{\pi }\over 2},
\ee
in order that the differential of (\ref{eqn:XY}) is linearly
transformed in the same way as (\ref{eqn:x_10}):
\be
(dY',dX')= (dY,dX) 
\left(
\begin{array}{cc}
a & b \\
c & d
\end{array}
\right)^{-1}
\label{eqn:dXdY'}.
\ee
This transformation preserves the metric \rf{mettor}. 

The transformation of the complex coordinates \rf{eqn:dXdY'} 
will cause a jump of 
the equation of a BPS membrane curve when the curve crosses a 
branch cut.
The equation of the membrane curve for a $(p,q)$ string (\ref{eqn:st}) can
be expressed as $qdX-pdY=0$ in the differential form.
Performing the transformation \rf{eqn:dXdY'}, 
we obtain the transformation of the $(p,q)$ charge as 
\be
\left(
\begin{array}{c}
-p' \\ q' 
\end{array}
\right)
= \left(
\begin{array}{cc}
a & b \\
c & d
\end{array}
\right) \left(
\begin{array}{c}
-p \\ q
\end{array}
\right).
\label{eqn:p'q'}
\ee
This transition of $(p,q)$ charges at a branch cut was essential 
in identifying the open string states corresponding to the gauge
bosons of the exceptional gauge groups in the references \cite{JOH,ZWI}.

In this short paper, we have discussed the holomorphic embedding of 
a membrane of M-theory to describe the BPS string configurations of the 
IIB string theory in the 7-brane backgrounds, and have investigated
some of their properties.
A general holomorphic embedding of a membrane describing a BPS string
configuration would be given by a holomorphic equation 
\be
f\left(\exp(R_BX),\exp(R_BY)\right)=0,
\ee
using the appropriate complex structure $X$ and $Y$ we have obtained
in \rf{eqn:XY}.
Among other things, 
the equations for 3-string junctions and string networks would
be given simply by substituting appropriately the variables
in the equations given in \cite{KROLEE}.

We have discussed infinitely stretched BPS string configurations 
in the 7-brane backgrounds in this papar. 
But, for further understanding of the BPS states and the 3-string
junction generating
process in the 7-brane backgrounds, an M-theory description of finitely 
stretched BPS string configurations ending on 7-branes would be desired.
 
\section*{Acknowledgements}

We wish to thank H.~Hata and Y.~Imamura for useful 
discussions and encouragement.
We are grateful to H.~Kunitomo and Y.~Imamura for the careful reading 
of the manuscript.


\begin{thebibliography}{99}

\bibitem{JOH} 
A.~Johansen, {\it A Comment on BPS States in F-theory in 8 Dimensions},
hep-th/9608186, Phys. Lett. {\bf B395} (1997) 36.

\bibitem{ZWI}
M.R.~Gaberdiel and B.~Zwiebach, {\it Exceptional groups from open strings},
hep-th/9709013.

\bibitem{HANWIT}
A.~Hanany and E.~Witten, {\it Type IIB superstrings, BPS monopoles, and 
three-dimensional gauge dynamics}, hep-th/9611230, 
Nucl. Phys. {\bf B492}, 152 (1997).

\bibitem{SHW}
J.H.~Schwarz, {\it Lectures on Superstring and M theory Dualities},
hep-th/9607201, Nucl. Phys. Proc. Suppl. {\bf 55B} (1997) 1.

\bibitem{DAS}
K.~Dasgupta and S.~Mukhi, {\it BPS Nature of 3-String Junctions},
hep-th/9711094. 

\bibitem{SENONE}
A.~Sen, {\it String Network}, hep-th/9711130.

\bibitem{REY}
S.-J~ Rey and J.-T.~Yee,  {\it BPS Dynamics of Triple (p,q) String
Junction}, hep-th/9711202.

\bibitem{KROLEE}
M.~Krogh and S.~Lee, {\it String Network from M-theory},
hep-th/9712050.

\bibitem{MAT}
Y.~Matsuo and K.~Okuyama, {\it BPS Condition of String Junction from 
M theory}, hep-th/9712070.

\bibitem{KOL}
B. Kol, {\it  5d Field Theories and M Theory}, hep-th/9705031. \CR
O. Aharony, A. Hanany and B. Kol, {\it Webs of (p,q) 5-branes, Five
Dimensional Field Theories and Grid Diagrams}, hep-th/9710116.

\bibitem{BECBEC}
K.~Becker, M.~Becker and A.~Strominger, {\it Fivebranes, Membranes and 
Non-Perturbative String Theory},
hep-th/9507158, Nucl. Phys. {\bf B456} (1995) 130.

\bibitem{GRESHA}
B.R.~Greene, A.~Shapere, C.~Vafa and S-T.~Yau, {\it Stringy Cosmic
Strings and Noncompact Calabi-Yau Manifolds}, Nucl. Phys. {\bf B337} (1990) 1.

\bibitem{GIBGRE}
G.W.~Gibbons, M.B.~Green and M.J.~Perry
{\it Instantons and seven-branes in type IIB superstring theory},
hep-th/9511080, Phys. Lett. {\bf B370} (1996) 37.

\bibitem{BERROO}
E.~Bergshoeff, M.~de~Roo, M.B.~Green, G.~Papadopoulos and
P.K.~Townsend, {\it Duality of Type II 7-branes and 8-branes},
hep-th/9601150, Nucl. Phys. {\bf B470} (1996) 113.

\bibitem{BHO} 
E.~Bergshoeff, C.M.~Hull and T.~Ortin, {\it Duality in the Type-II
 Superstring Effective Action},
hep-th/9504081, Nucl. Phys. {\bf B451} (1995) 547.

\bibitem{WIT}
E.~Witten, {\it String Theory Dynamics In Various Dimensions},
hep-th/9503124, Nucl. Phys. {\bf B443} (1995) 85.

\bibitem{ASP} 
P.S.~Aspinwall, {\it K3 Surfaces and String Duality}, hep-th/9611137.

\bibitem{SENTWO}
A.~Sen, {\it BPS States on a Three Brane Probe}, hep-th/9608005,
Phys. Rev. {\bf D55} (1997) 2501.


\end{thebibliography}
\end{document}